\documentclass[apl,twocolumn,showpacs]{revtex4}

\usepackage{graphicx}
\usepackage[dvips]{color}

\newcommand{\gapprox}{{\scriptscriptstyle\stackrel{>}{\sim}}}
\newcommand{\lapprox}{{\scriptscriptstyle\stackrel{<}{\sim}}}

\begin{document}

\title{Imaging of Thermal Domains in ultrathin NbN films for Hot Electron
Bolometers}

\author{D. Doenitz, R. Kleiner, D. Koelle}
\affiliation{ Universit\"{a}t T\"{u}bingen, Physikalisches Institut --
Experimentalphysik II, Auf der Morgenstelle 14, D-72076 T\"{u}bingen, Germany}

\author{T. Scherer, K. F. Schuster}
\affiliation{Institute de Radioastronomie Millimetrique, IRAM, 300
Rue de la Piscine, 38406 St Martin d'H\`eres, France}

\date{May 26, 2007}

\begin{abstract}
We present low-temperature scanning electron microscopy (LTSEM)
investigations of superconducting microbridges made from ultrathin
NbN films as used for hot electron bolometers.
LTSEM probes the thermal structure within the microbridges under
various dc current bias conditions, either via electron-beam-induced
generation of an unstable hotspot, or via the beam-induced growth of
a stable hotspot.
Such measurements reveal inhomogeneities on a micron scale, which may
be due to spatial variations in the NbN film or film-interface
properties.
Comparison with model calculations for the stable hotspot regime
confirm the basic features of common hot spot models.
\end{abstract}


\maketitle

Ultrathin NbN films have recently gained much attention for use in
THz hot electron bolometer (HEB) mixers and fast photon counting
detectors (see e.g. \cite{Gousev94,Kawamura96,Lehnert98,Goltsman01}).
NbN HEB mixers are short microbridges (typical length and width of
order 1\,$\mu$m and a few nm thickness) with contacts to antenna and
dc leads.
The dc current-voltage curve (IVC) is characteristic for microbridges
with a supercurrent up to a critical current $I_c$ and a sudden jump
to a hysteretic resistive behavior.
Depending on film thickness, quality and geometry, typical
resistances for this branch are around 100--1000\,$\Omega$.
When current biased on the resistive branch, the bridge stays
resistive below $I_c$, down to a distinct voltage from where the
device jumps back to the zero-voltage state at the return current
$I_r$.
When the bridge is biased at the lower stable voltage carrying region
it serves as a sensitive power detector for far-infrared (FIR)
radiation.
The basic principle of heterodyne mixing with such a device is the
beating of an incoming signal with an auxiliary local oscillator (LO)
signal, producing a signal at the intermediate frequency (IF), the
difference between the frequency of the input signal and the LO
frequency.
The resulting IF signal lays in the 0--4\,GHz range and can therefore
be amplified with commercial semiconductor low-noise amplifiers.
We note that for sufficient LO power, the IVC becomes non-hysteretic,
and the HEB mixer is typically voltage biased.

NbN microbridge HEB mixers offer several very important advantages
over other mixer devices in the THz range.
Their small size allows to pump the mixer with very little LO power;
this is an important advantage, as tunable LO sources above 500\,GHz
have usually very little power output.
Another advantage is due to the purely resistive nature of the
device, which makes high-frequency impedance matching very easy.
Because the detection is based on heating, the detection principle is
not limited in frequency by specific superconducting properties, such
as the energy gap of NbN, as in the case of SIS mixers.
For frequencies between 1.2 and 10\,THz, HEB mixers are therefore
offering currently the lowest noise.
However, current NbN HEB mixers suffer from a relatively low IF
bandwidth.
This is due to a limited detection speed, which is often found to be
below 2\,GHz.
For THz applications, IF bandwidths up to 10\,GHz are however
desirable in radio astronomy.
The detailed physics of the involved time constants, which ultimately
limit the IF bandwidth, is therefore a subject of ongoing research in
many different laboratories.

While first results for these applications are promising
\cite{Cherednichenko02,Wiedner06},
modelling of the devices is quite complex.
As a consequence, the impact of fabrication and specific material
parameters on device performance remains unclear, and optimization is
slow and based on trial and error.
Beyond first lumped element models \cite{Karasik96}, more elaborated
models have been proposed \cite{WilmsFloet99,Merkel00}, which
describe the physics of the microbridges in a spatially resolved
manner.
Most of the numerical and analytical spatially distributed models are
based on assumptions first made for superconducting bridges by
Skocpol, Beasley and Tinkham \cite{Skocpol74}.
In this approach the resistive behavior of a superconducting
microbridge is modelled through the thermal equilibrium between a
normal conducting and therefore dissipating hot spot and the
substrate.
The lateral thermal conduction and the cooling to the substrate
determine the detailed shape of the hot spot.
The general solution within this analytical frame work is a
symmetrical hot spot geometry centered between the contacts.
Although some convincing results, such as modelling of rf-pumped IVCs
and bias dependent conversion gain curves could be obtained for the
current NbN HEBs within the existing distributed models, the
existence of a single centered hot spot geometry for these devices
has not been directly shown so far.

Over the last two decades, low-temperature scanning electron
microscopy (LTSEM) has been used to provide local information (on a
micron scale) on various properties of superconducting thin films and
Josephson junctions, such as the spatial distribution of the
transition temperature $T_c$ and critical current density $j_c$, or
on Josephson vortices in long
junctions\cite{Clem80,Huebener84,Gross94}, and on Abrikosov
vortices\cite{Doenitz04,Straub01} and supercurrent
distribution\cite{Doenitz06} in SQUID washers.
Very early, LTSEM has been applied to proof the concept of hot spot
formation in long and thick superconducting bridges
\cite{Eichele81,Eichele82,Eichele83}.
In this paper we show that LTSEM can also be applied to investigate
the thermal structures of microbridges from ultrathin NbN films with
much smaller geometries, approaching those of HEBs.

\begin{figure}[t]
\noindent\center\includegraphics[width=8.5cm]{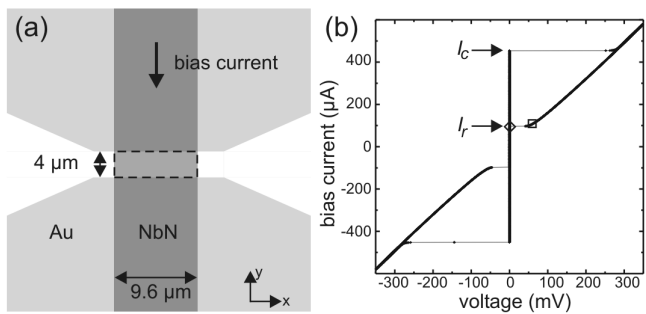}
\hfill\caption{ (a) NbN HEB layout:
$W=9.6\,\mu$m wide NbN strip, partially covered by Au pads with
$L=4\,\mu$m separation.
(b) IV characteristic measured at $T\approx 5\,$K in the LTSEM.
Open diamond and circle show bias points for LTSEM images shown in
Fig.~2(b) and Fig.~3, respectively} \label{Layout}
\end{figure}

The samples which we investigated were fabricated as follows:
Thin NbN films were deposited from a 4 inch Nb target by 13.56~MHz
(240\,W) rf magnetron sputtering on a 2 inch fused quartz substrate
at room temperature in a 0.852~Pa N$_2$/Ar/CH$_4$ atmosphere (gas
flow: 2.7, 46 and 0.6~sccm, respectively).
This process yields typically $T_c = 11\,$K for $t = 5\,$nm thick
films and $T_c = 15\,$K for $t > 20\,$nm.
The NbN films were patterned by reactive ion etching in a
CF$_4$/O$_2$ mixture to form long microbridges of width $W$.
Subsequently, 50\,nm thick Au contact pads were formed on top of the
NbN bridges, with a separation, which defines the length $L$ of the
active region of the NbN HEB, as shown in Fig.~\ref{Layout}(a).
Typical device geometries vary from $L \times W = 0.5\,\mu{\rm m}
\times 4\,\mu$m to $5\,\mu{\rm m} \times 10\,\mu$m.
We present results from a device with $t=5\,$nm, $L=4\,\mu$m and
$W=9.6\,\mu$m, which shows a clearly hysteretic IVC at $T\approx
5\,$K [c.f.~Fig.~\ref{Layout}(b)].
We note that all data presented here have been obtained with current
bias.

For imaging by LTSEM, the sample was mounted on a liquid He cooled
stage and operated at a temperature $T\approx 5\,$K.
The local perturbation by the focused electron beam (e-beam) induces
an increase in temperature $\delta T(x,y)$ on the sample surface in
the $(x,y)$ plane, on a length scale of approximately $1\,\mu$m,
which determines the spatial resolution of this imaging technique,
and with a maximum local increase in temperature $\Delta T(x_0,y_0)$
of a fraction of 1\,K, centered on the beam spot position $(x_0,y_0)$
on the sample surface \cite{Clem80}.
For modelling the local perturbation we have used a combination of
Monte-Carlo calculations for the beam energy deposition and finite
element modelling (FEM) of the corresponding thermal plume.
However, simplified semi-analytical models turned out to give very
similar results \cite{Gross94}.
The local change in $T$ may change global properties of the bridge,
e.g.~the voltage $V$ across the current-biased bridge.
This voltage change $\delta V$ depends on the e-beam position
$(x_0,y_0)$ and thus can be recorded to obtain a $\delta
V(x_0,y_0)$-image.
To improve the signal-to-noise ratio, we use a beam-blanking unit
operating at 5\,kHz and detect $\delta V$ with a lock-in amplifier.

In order to create a voltage drop along the microbridge, a continuous
domain (hot spot) of normal conducting film across the bridge is
required.
Such a domain can be generated either by resistive heating due to the
bias current $I_b>I_r$ or, for lower bias currents, by the combined
effect of e-beam and current heating.
It is therefore possible to distinguish two different experiments:\\
{\em (A) e-beam-induced generation of (unstable) hot spot:}
If biased below $I_r$, the device may switch to a resistive state
upon e-beam irradiation due to generation of an unstable hotspot
(i.e.~the hotspot disappears when the e-beam is turned off).
The lock-in detected voltage signal corresponds to the voltage
induced by the (unstable) hotspot.\\
{\em (B) e-beam induced growth of (stable) hotspot:}
If biased slightly above $I_r$ in the resistive state, a stable
hotspot is generated, which size may be altered upon e-beam
irradiation.
The lock-in detected voltage signal corresponds to the beam spot
position-dependent small change in voltage drop due to the
beam-induced extension of the otherwise self-sustained hotspot.

\begin{figure}[b]
\noindent \center\includegraphics[width=8.5cm]{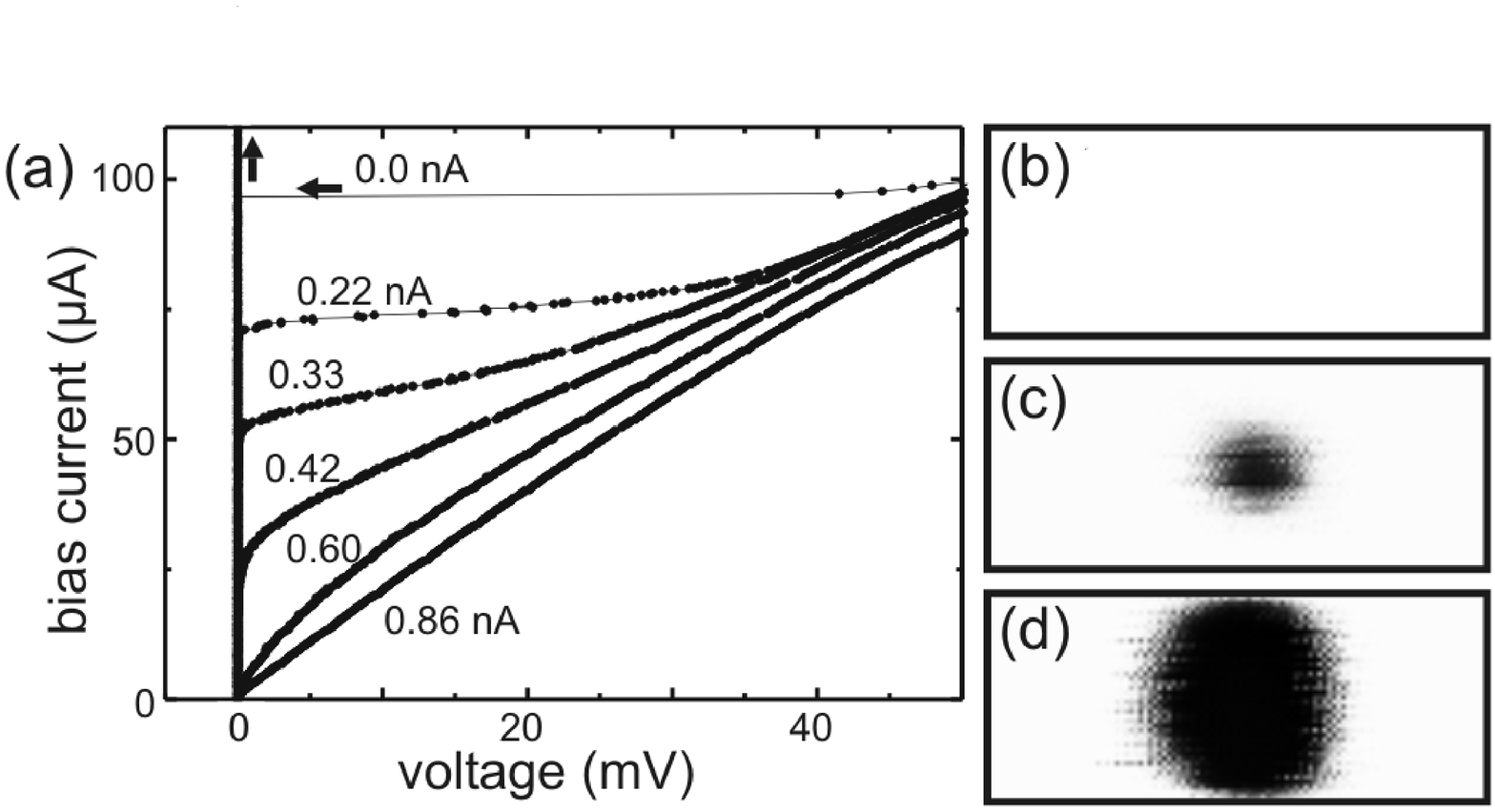}
\hfill\caption{ (a) Non-hysteretic IVCs for e-beam irradiation on the
center of the device ($I_{el}$ from 0.22 to 0.86\,nA;
$U_{el}=20\,$kV); unirradiated hysteretic IVC is shown for comparison
(arrows indicate sweep direction of $I_b$).
(b-d) LTSEM voltage images ($I_b\approx 95\,\mu$A, i.e.~just below
$I_r$, $U_{el}=10\,$kV) for different $I_{el}$: (b) 0.24\,nA, (c)
0.26\,nA, (d) 0.28\,nA.
The rectangles mark the boundaries of the sample [c.f.~dashed box in
Fig.\ref{Layout}(a)].} \label{induzHS}
\end{figure}

We first discuss imaging mode (A).
Figure \ref{induzHS}(a) shows IVCs recorded during e-beam irradiation
(for various values of the beam current $I_{el}$ and fixed beam
voltage $U_{el}=20\,$kV) at a fixed position on the center of the
device, and for comparison, the IVC without irradiation.
With increasing $I_{el}$ the critical current is reduced, while the
normal resistance remains almost unchanged.
Under sufficiently strong e-beam irradiation ($I_{el}\gapprox
0.2\,$nA) the hysteresis in the IVC vanishes.
The general shape of the IVCs with e-beam irradiation is surprisingly
similar to IVCs of rf pumped devices\cite{WilmsFloet99}.
This is a strong indication that a general mechanism, which is likely
to be of thermal nature, describes the physics of the IVCs for very
different types of energy input.

Figure \ref{induzHS}(b-d) shows LTSEM voltage images recorded at
slightly sub-critical current bias ($I_b\lapprox I_r$) for fixed
$U_b$ and different values of $I_{el}$.
These images reflect the local sensitivity to e-beam irradiation for
triggering of an unstable hot spot.
For $I_{el}=0.24\,$nA (b), no hot spot is induced at all.
With an increase to $I_{el}=0.26\,$nA (c) a small spot on the image
appears: only when the e-beam is applied within this spot a voltage
is induced.
As expected from thermal and electrical symmetry considerations, the
spot is centered with respect to the $x$-axis.
It is also centered with respect to the $y$-axis (along the bridge),
which can be easily explained by the additional cooling effect
provided by the gold pads.
With further increase of $I_{el}$ to 0.28\,nA, the spot increases
rapidly [see Fig.~\ref{induzHS}(d)].
An analysis of the shape of the spots leads to the estimate of the
spatial resolution of $\approx 0.4\,\mu$m for this type of
measurement.

\begin{figure}[t]
\noindent \center\includegraphics[width=8.5cm]{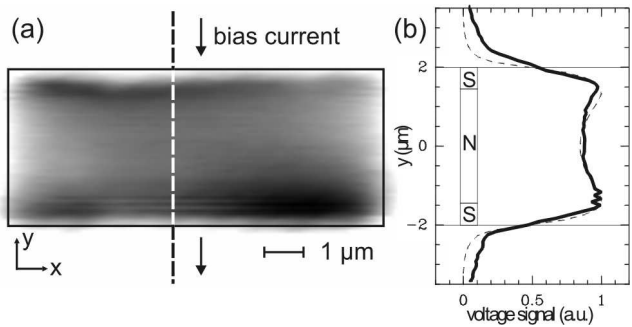}
\hfill\caption{ (a) LTSEM voltage image at $I_b=110\,\mu$A (stable
hot spot condition); $I_{el}=10\,$pA, $U_{el}=5\,$kV.
Dark areas indicate high voltage responsivity to e-beam irradiation.
(b) Line-scan of LTSEM voltage signal (solid curve) along the dashed
line in (a) and comparison with model calculation using 2-dimensional
FEM techniques (dashed curve). Inset indicates the transition between
superconducting (S) and normal (N) region at $y=\pm 1.4\,\mu$m.}
\label{c44_2}
\end{figure}

For imaging mode (B), ($I_b>I_r$), i.e.~in the area of a stable bias
current driven hot spot, the voltage across the sample increases
slightly under irradiation due to the following effect:
The beam-induced increase in temperature causes the hot spot to grow;
thus an increased region of the NbN film becomes resistive, and the
normal resistance $R$ is increased.
This mechanism is most effective for an e-beam position close to the
superconducting/normal conducting (S/N) boundary.
The result is a characteristic double ridge picture, with the maximum
of the signal indicating the location of these lateral boundaries, as
shown in Fig.~\ref{c44_2}(a).
However, the LTSEM signal also reveals a clear variation along the
S/N boundaries.
This may be due to inhomogeneities in the film quality such as
thickness or composition, or due to local variations in the thermal
coupling to the substrate or contact pads.
The detected inhomogeneities will result in an asymmetric current
density and may adversely affect device performance.

We modelled the device under the stable hotspot bias conditions as in
imaging mode (B) by a 2-dimensional finite element calculation,
taking into account the heat transfer to the substrate as balanced by
the heat production by the bias current dissipation.
The e-beam induced change of the substrate surface temperature was
modelled using Monte-Carlo techniques to compute the energy input and
a finite element model in the symmetry plane of the e-beam axis to
derive the temperature resulting from this energy input.
For the parameters ($I_b$, $I_{el}$, $U_{el}$) and device geometry
used in the experiment [c.f.~Fig.\ref{c44_2}], the model predicts a
superconducting (S)/normal conducting (N) transition at a position of
$y=\pm 1.4\,\mu$m, as indicated in Fig.\ref{c44_2}(b).
The derived voltage signal $\delta V(y)$ for a linescan along the
$y$-direction [c.f.~Fig.~\ref{c44_2}(b)] reproduces nicely the
characteristic double peaked shape as observed experimentally, and
matches also very well the general form of the measured curve.

In conclusion, we demonstrated that LTSEM is a useful tool to
investigate thermal domain (hotspot) formation in ultrathin
superconducting films.
Our experiments show that for current bias below $I_r$, e-beam
irradiation induces a hotspot, depending on deposited beam energy,
beam spot position and sample inhomogeneity.
For biasing above $I_r$, i.~e.~in the regime of a stable hotspot,
comparison of measurements with 2-dimensional modelling shows that
the classical hotspot model does apply.
Our investigations also show that inhomogeneities on a micron scale
can be detected.
Further investigations are required to clarify the nature of such
inhomogeneities and their impact on device performance.
Furthermore, it will be interesting to include RF pumping to
investigate the hotspot formation under typical working conditions of
superconducting hot electron bolometers.

D.D.~gratefully acknowledges support from the Evangelisches
Studienwerk e.V. Villigst. This work was supported by the EU FP6
Program RADIONET (AMSTAR).


\end{document}